# KAN-Enhanced Contrastive Learning Accelerating Crystal Structure Identification from XRD Patterns

Chenlei Xu[a,1], Tianhao Su[a,1], Jie Xiong[a,d]*, Yue Wu[a], Shuya Dong[a], Tian Jiang[a],

Mengwei He[b,c]*, Shuai Chen[a,d], Tong-Yi Zhang[a,e]*

[a] Materials Genome Institute, Shanghai University, Shanghai 200444 China

[b] Australian Centre for Microscopy and Microanalysis, School of Aerospace, Mechanical and Mechatronic Engineering, The University of Sydney, Sydney, New South Wales, 2006, Australia

[c] School of Computer Science, The University of Sydney, Sydney, New South Wales, 2006, Australia

[d] State Key Laboratory of Materials for Advanced Nuclear Energy, Shanghai University, Shanghai 20044, PR China

[e] Guangzhou Municipal Key Laboratory of Materials Informatics, Hong Kong University of Science and Technology (Guangzhou), Guangzhou 511400, China

[1] These authors contributed equally to this work and should be considered co-first authors.

E-mail: xiongjie@shu.edu.cn (J. Xiong), mengwei.he@sydney.edu.au (M. He), zhangty@shu.edu.cn (T.-Y. Zhang)

**Abstract** Accurate determination of crystal structures is central to materials science, underpinning the understanding of composition–structure–property relationships and the discovery of new materials. Powder X-ray diffraction (XRD) is a key technique in this pursuit due to its versatility and reliability, yet current analysis pipelines still rely heavily on expert knowledge and slow iterative fitting, limiting their scalability in high-throughput and autonomous settings. Here we introduce a physics-guided contrastive learning framework, XRD–Crystal Contrastive Pretraining (XCCP), which aligns powder diffraction patterns with candidate crystal structures in a shared embedding space to enable efficient structure retrieval and symmetry recognition. The XRD encoder employs a dual-expert design with a Kolmogorov–Arnold Network projection head: one branch emphasizes low-angle reflections reflecting long-range order, while the other captures dense high-angle peaks shaped by symmetry. Coupled with a crystal graph encoder, contrastive pretraining yields physically grounded representations. XCCP demonstrates strong performance across tasks, with structure retrieval reaching 88.98% and space group identification attains 93.39% accuracy. The framework further generalizes to compositionally similar multi-principal element alloys and demonstrates zero-shot transfer to experimental patterns. Together, these results establish XCCP as a robust, interpretable, and scalable approach that offers a new paradigm for PXRD analysis, facilitating high-throughput screening, rapid structural validation, and integration into autonomous laboratories.

**Keywords**: Powder X-ray diffraction; Crystal structure determination; Machine learning; High-throughput materials screening; Autonomous laboratories.

# 1. Introduction

Powder X-ray diffraction (XRD) is a foundational analytical technique for characterizing crystalline materials, providing diffraction signatures that reveal key crystallographic parameters such as lattice constants, space groups, and phase compositions[1-5]. These capabilities have made XRD indispensable for microstructural characterization across materials science and engineering. However, conventional analysis workflows face notable limitations. Manual peak assignment based on the Bragg equation followed by database matching remains standard practice, requiring substantial crystallographic expertise, hindering operational efficiency, and becoming particularly challenging when diffraction peak overlaps[6-8]. Rietveld refinement, although widely used, depends on accurate initial assumptions and reliable reference powder diffraction file (PDF) cards[9, 10], and its iterative nature constrains rapid characterization in high-throughput materials discovery[11, 12].

Recent advances in data-driven methods have shown potential to address these challenges[13-18]. Deep neural networks can capture complex, nonlinear relationships inherent in high-dimensional diffraction data. Convolutional neural networks (CNNs) can resolve overlapping XRD peaks [19, 20] and attention mechanisms have shown strong capabilities in capturing polycrystalline textures[21-24]. These models have demonstrated superior performance in classification tasks, often exceeding traditional methods[25-32]. For example, Lee et al.[28] trained CNNs on approximated 1.7 million synthetic XRD patterns generated from the Sr-Li-Al-O quaternary and achieved near-perfect identification in multi-phase settings. Cao et al.[33] developed a XRD simulation method that incorporates comprehensive physical interactions, resulting in a high-fidelity database including 4,065,346 simulated powder XRD patterns, representing 119,569 unique crystal structures under 33 simulated conditions that reflect real-world variations. Salgado et al.[34] reported a non-pooling CNN trained on 170,000 inorganic crystal structures that achieved 67% accuracy for crystal systems and only 36% for space groups on 2,253 test samples. Nevertheless, most existing efforts treat XRD primarily as a symmetry-assignment task, whereas in practice the central challenge is retrieval—matching an observed pattern to candidate entries in a reference database[35-37].

Cross-modal contrastive learning offers a natural framework for this problem, as it learns shared representations that bring each diffraction pattern closer to its corresponding structure. Xie et al.[38] proposed a contrastive learning system that couples a mass spectrometry spectral encoder with a molecular structural encoder and reports notable gains in retrieval. A similar strategy in crystallography can align diffraction patterns with candidate structures and enable fast, accurate retrieval. The success of such a framework depends on encoders for both crystals and XRD signals. Crystal encoding has been studied extensively, and several effective graph neural networks (GNNs) are available[19, 39-42]. Representative examples include the crystal graph convolutional neural network (CGCNN)[19], the atomistic line GNN (ALIGNN)[40], the GNN with three-body interactions (M3GNet)[39] and related variants that process atomistic graphs with learned message passing.

Existing architectures are largely inherited from symmetry-prediction tasks, with limited incorporation of diffraction-specific inductive biases. Addressing this gap is critical for improving sample efficiency and fidelity in retrieval. Kolmogorov-Arnold Networks (KAN)[43-45] provide a promising direction for representing the highly nonlinear and fluctuating character of XRD signals. Unlike conventional multilayer perceptron (MLP) models that rely on fixed activation functions, KANs adopt learnable spline-based activation functions that enable adaptive nonlinear transformations, thereby improving the efficiency when representing complex mappings[44-48]. Jia et al.[46] combined KAN with bidirectional sequence modeling for lithium-ion battery degradation and achieved superior accuracy under capacity self-recovery and drift. Wu et al.[49] proposed a contrastive pretraining to align crystal structures with their physical properties and showed that KAN projection heads outperformed MLP heads in accuracy and convergence. In crystallography, such adaptive can enhance representation learning for diffraction signatures and can support robust cross-modal alignment.

Guided by these insights, we propose a unified XRD-Crystal Contrastive Pretrained (XCCP) framework for automated crystal structure analysis from powder XRD patterns. The framework couples a dual-expert XRD encoder, which attends separately to small-angle reflections and dense wide-angle peaks, with a KAN projection head, and aligns it to a crystal-graph encoder through contrastive training. The aligned embedding enables direct structure retrieval from a query powder pattern and supports accurate space group inference when required. The framework achieves efficient retrieval, and elemental pre-screening further narrows candidate pools for practical deployment. These advances establish XCCP as a physically grounded and scalable approach for PXRD analysis, supporting high throughput screening, rapid validation, and integration with autonomous laboratory platforms.

## 2. Methods

**2.1. Data Preparation**

The dataset contains 155,003 crystallographic information files (CIFs) from the Materials Project [50] database, spanning a wide range of crystal structures and chemical compositions. **Fig. 1a** summarizes the distribution across the seven crystal systems. **Fig. 1b** reports the ten most frequent space groups among 220 reported in the database. The histogram of space groups displays a pronounced long tail, which can bias classification toward majority space groups. This imbalance reflects the natural prevalence of crystalline symmetries and is therefore retained. No filtering or artificial rebalancing was applied so that class proportions remain representative of real materials problems.

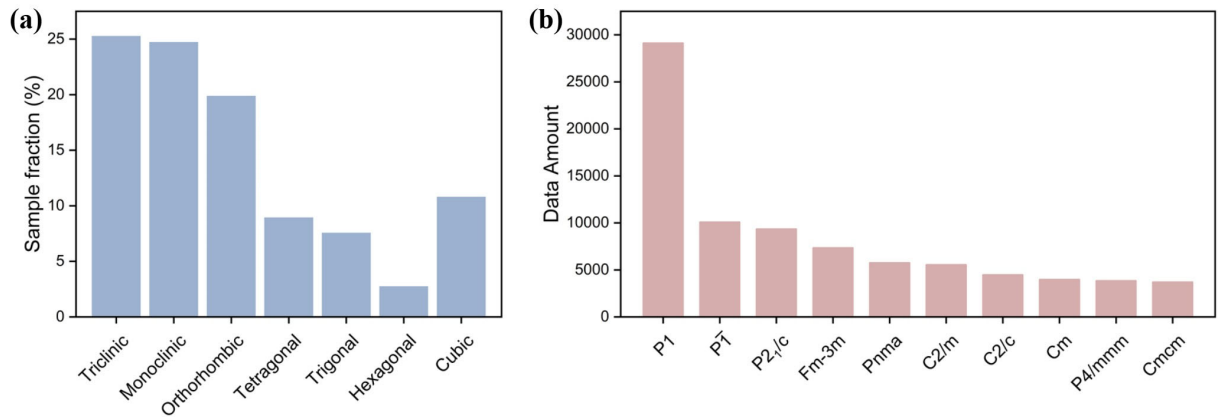

**Fig. 1**. **Crystallographic data analysis.** (a) Statistical distribution of the seven crystal systems present in the dataset. (b) Frequency distribution of the ten most prevalent space groups among 220 labels.

Powder XRD patterns were simulated from CIFs using *PyXtal*[51], yielding 155,003 synthetic diffraction profiles. For each structure, crystallographic planes $hkl$ were identified and their interplanar spacings calculated from reciprocal lattice geometry. Bragg's law is enforced to identify planes that satisfy the angular and wavelength conditions,

$$n\lambda = 2d_{hkl}sin\theta \qquad (1)$$

where $\lambda$ is the X-ray wavelength and $\theta$ is the diffraction angle. For each eligible plane, the relative diffraction intensity $I_{hkl}$ is evaluated from the structure factor $F_{hkl}$ and the Lorentz-polarization corrections factor $L_p$,

$$I_{hkl} \propto |F_{hkl}|^2 \cdot L_p \qquad (2)$$

with

$$F_{hkl} = \sum_j f_j exp[2\pi i(hx_j + ky_j + lz_j)] \qquad (3a)$$

$$L_p = \frac{1 + cos^2 2\theta}{sin^2 \theta cos \theta} \qquad (3b)$$

Here $f_j$ is the atomic scattering factor and $(x_j, y_j, z_j)$ are the atomic fractional coordinates. Each pattern was intensity-normalized by the maximal peak intensity before model training to ensure stable optimization.

Most simulated powder XRD studies [33, 52] only consider the range $10° \leq 2\theta \leq 80°$. The present work follows this convention and refer to it as the wide-angle (WA) XRD profile. XRD patterns at $2\theta < 10°$ were also simulated here, denoted as small-angle (SA) XRD. The statistical analysis (see **Supplementary Materials**) of simulated patterns shows that the dominate signal lies at WA region, while non-negligible reflections remain in the SA range. These SA peaks correspond to large *d*-spacings from Bragg's law. They often report interlayer distances, superlattice ordering, guest induced expansion, and other long period motifs. Such simulation can improve discrimination among structures with similar high angle fingerprints.

The final dataset, termed MP-SXRD, thus contains 155,003 paired crystal structures and simulated diffraction patterns profiles with $0° < 2\theta \leq 80°$. It was randomly partitioned into training (70%), validation (10%), and test (20%) subsets. Stratified sampling preserved the distribution of space group across splits so that evaluation remains faithful to the long-tailed character of the source set.

**2.2 XRD-Crystal Contrastive Pretraining and Employment**

The proposed framework aligns powder XRD patterns and crystal structures in a shared latent space, as illustrated in **Fig 2**. Two modality specific encoders are trained jointly to produce comparable embeddings that support cross-modal retrieval and symmetry inference. The crystal encoder is a modified CGCNN that is widely used and easy to reproduce. Each structure is represented as a graph of atoms and bonds. Message passing aggregated local chemical environments, and a global pooling layer yields a 64-dimensional crystal embedding $v_c$.

Powder diffraction patterns are encoded with a dual-expert network incorporating a Kolmogorov–Arnold Network (KAN) projection head (DEN-KAN). The DEN-KAN encoder follows the SAXRD and WAXRD ranges defined in section 2.1. Two parallel branches process the two measurement ranges in tandem, and the KAN-based projection head fuses their outputs. Each branch adopts a ResNet backbone with residual connections and hierarchical feature extraction capabilities[53]. The two branches use the same residual blocks and ReLU activation functions. The WAXRD pathway applies max pooling with stride two and 1×3 kernels after each residual layer to enlarge the receptive field for closely spaced peaks. The outputs are concatenated and passed to the KAN-based projection head, producing a 64-dimensional XRD embedding $v_{xrd}$.

For settings where small-angle coverage is unavailable, a single-path variant that ingests only the conventional WAXRD range starting at $2\theta = 10°$ was also developed. This variant keeps the WAXRD branch and the same KAN-based projection head (referred to as WA-KAN), with implementation details provided in **Supplementary Materials**. Both DEN-KAN and WA-KAN encoders are trained end to end with the same contrastive loss so that crystal and diffraction embeddings remain comparable.

Training maximizes the agreement and reduces the disagreement for XRD pattern-structure pairs with a symmetric InfoNCE loss $\mathcal{L}$,

$$\mathcal{L}(\theta_{xrd}, \theta_c) = \frac{1}{2}[\mathbb{E}_{(i,j)}[-\log \frac{e^{s(\mathbf{v}_{xrd}^i, \mathbf{v}_c^j)/\tau}}{\sum_{k=1}^{N} e^{s(\mathbf{v}_{xrd}^i, \mathbf{v}_c^k)/\tau}}] + \mathbb{E}_{(i,j)}[-\log \frac{e^{s(\mathbf{v}_c^j, \mathbf{v}_{xrd}^i)/\tau}}{\sum_{k=1}^{N} e^{s(\mathbf{v}_c^j, \mathbf{v}_{xrd}^k)/\tau}}]] \quad (4)$$

where $s(\cdot, \cdot)$ is cosine similarity, hyperparameter $\tau = 0.07$ is used here to appropriately control the similarity distribution sharpness, and $N = 128$ is the batch size. We employ a MultiStepLR learning rate schedule and batch normalization for stable training and regularization. Trainable parameters of both encoders $\theta_{xrd}$ and $\theta_c$ are optimized jointly.

Inference follows a two-stage pipeline. Candidate structures are first filtered by chemical composition. Cosine similarity, between the XRD-encoder embedding of the query pattern and the CGCNN embeddings of the filtered structures, then produces a ranked prediction list with the top-$k$ ($k = 1,2,\cdots,$) candidates for the query XRD pattern. Cross-modal retrieval performance is evaluated by top-$k$ matching accuracy, defined as the probability that the correct structure appears among the first $k$ candidates retrieved for a given XRD query.

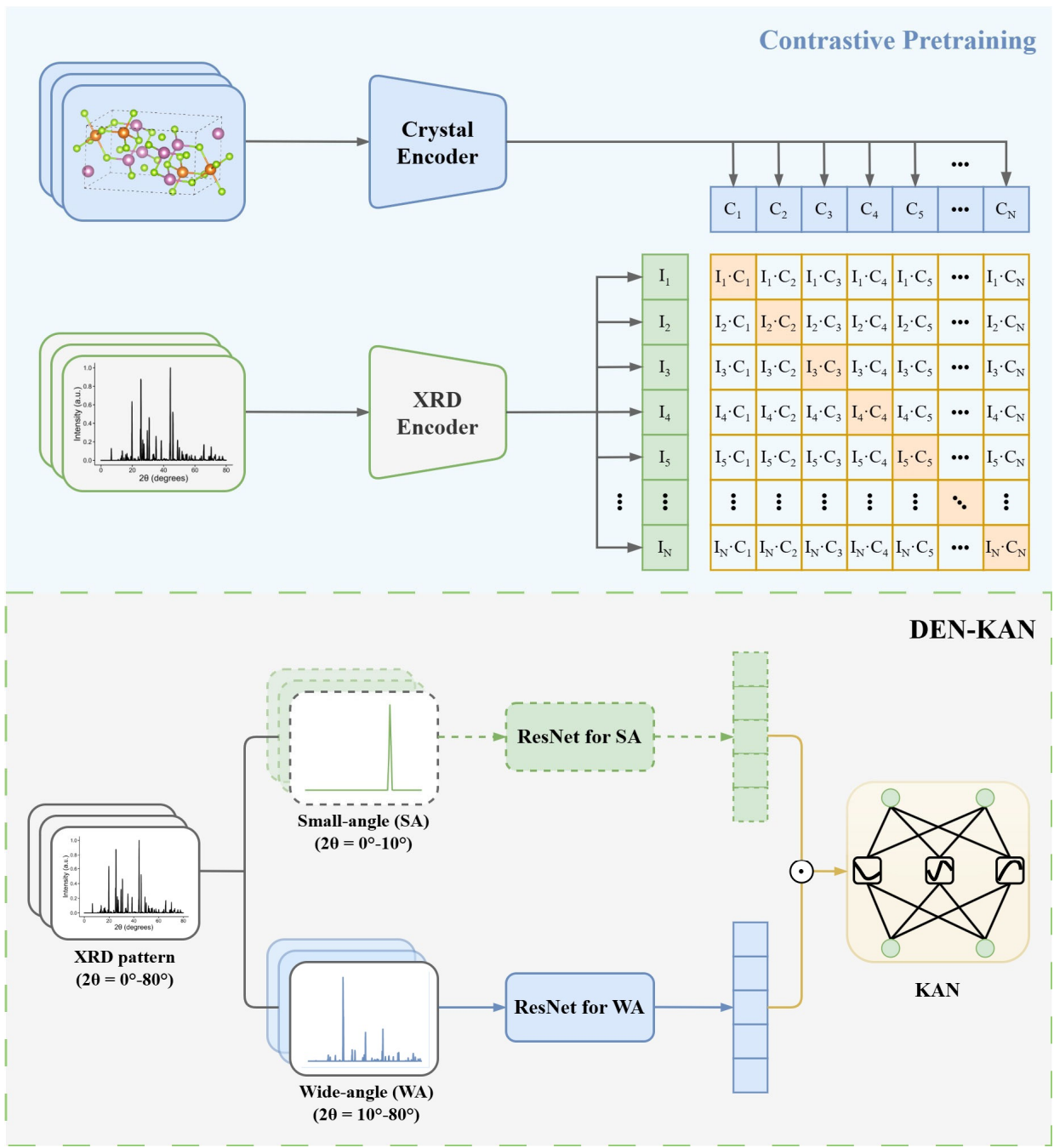

**Fig. 2**. **XCCP framework for crystal–XRD alignment and retrieval.** (a) Contrastive training aligns powder XRD patterns with their crystal structures. A CGCNN-based crystal encoder produces structure embeddings. The dual-expert XRD encoder (DEN-KAN) encodes the diffraction pattern and can handle small-angle (SAXRD) and wide-angle (WAXRD) inputs, and a KAN head fuses the branch features to yield the XRD embedding. When small-angle data are unavailable, only the wide-angle branch is used.

# 3. Results

## 3.1 Performance of XCCP in structure retrieval

The XCCP framework retrieves crystal structures from powder XRD patterns with high reliability. In **Fig. 3a**, XCCP equipped with the proposed DEN-KAN encoder achieves a 46.42% top-1 accuracy on the test set without elemental priors. Incorporating elemental information raises top-1 accuracy to 88.98%, which exceeds the 67.8% reported for *Jade* software under the same protocol [24]. With elemental filtering, the retrieval accuracies of top-3 and top-5 increase to 97.56% and 98.82%, respectively. In routine workflows, this ensures that the correct structure appears within the first three candidates for almost all queries, enabling rapid short-list inspection.

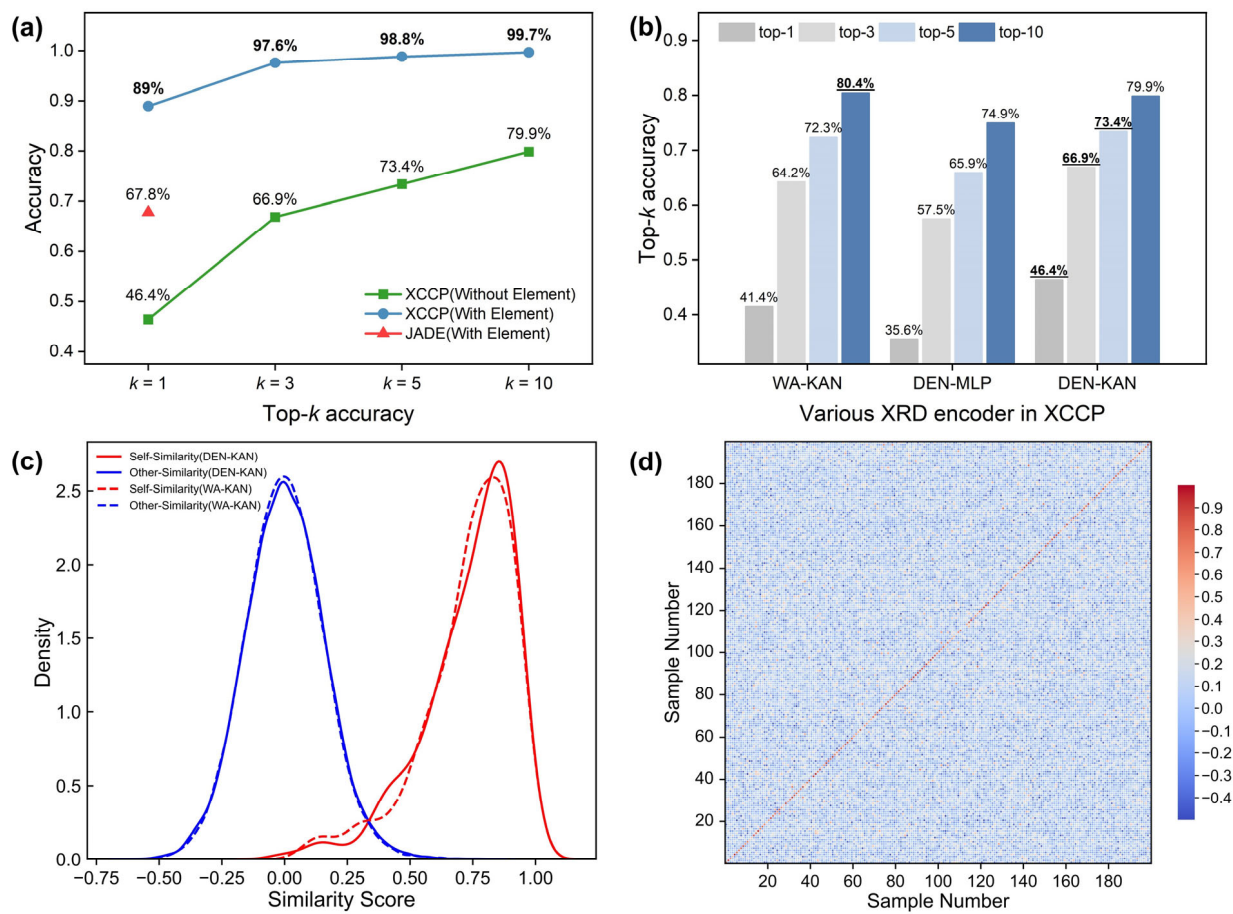

**Fig. 3.** **Retrieval accuracy and similarity analysis**. (a) Top-$k$ retrieval accuracy of the XCCP framework with and without elemental filtering. (b) Ablations across different XRD encoders. (c) self-similarity and other-similarity distributions. (d) Cosine-similarity heat map for representative batches.

Architecture-level ablations without elemental filtering clarify the sources of gain, as shown in **Fig. 3b**. A single-path variant that retains only the WAXRD branch with a KAN head (WA-KAN) yields 41.4% top-1, 64.2% top-3, 72.3% top-5, and 80.4% top-10 accuracy. Introducing SAXRD increases information content and sharpens early-rank precision. The DEN-KAN surpasses WA-KAN at small $k$, while a mild trade-off appears at larger $k$ ($k = 10$), as fusion steepens the similarity landscape and concentrates scores on a few strong candidates. Near-miss structures that remain above the cutoff under WA-KAN can drop below it after fusion. The net effect is higher precision at small $k$ with a modest reduction in recall at large $k$. With elemental constraints, the top-3 accuracy of DEN-KAN is already very high and supports confident identification.

The projection head has a much stronger impact on retrieval than the backbone. Replacing the KAN head with a conventional MLP while keeping the dual-expert design (DEN-MLP) gives the weakest performance across ranks. Even without SAXRD information, WA-KAN outperforms DEN-MLP by about 6% top-1 accuracy. The KAN head aligns well with the diffraction representations and supplies the decisive improvement for top-$k$ retrieval in this setting. Its adaptive spline activations [45, 48] capture peak shapes and background trends within a flexible expansion, while local support increases tolerance to small 2θ shifts and missing weak reflections. Stable gradients under symmetric InfoNCE further enhance retrieval reliability.

Similarity analysis provides additional insights. Self-similarity denotes the cosine similarity between an XRD pattern and its true crystal structures, while other-similarity refers to mismatched XRD-structure pairs. As shown in **Fig. 3c**, both DEN-KAN or WA-KAN separate these two distributions, yet DEN-KAN shifts the self-similarity peak to a higher value and narrows its spread. The tail of the other-similarity curve intrudes less into the self-similarity region, which reduces false positives and explains the higher precision at small $k$. WA-KAN exhibits broader self-similarity tails and therefore higher recall at large $k$. The cosine-similarity heat map for XCCP equipped with DEN-KAN in **Fig. 3d** shows strong diagonal dominance, confirming that each query aligns most strongly with its ground-truth structure, while off-diagonal values remain low. These results validate the architectural choice and confirm practical utility for high-throughput analysis.

## 3.2 Interpretability and physical priors of the DEN-KAN encoder

Generalization across symmetry classes benefits from XRD-encoders that reflect how diffraction features form. The dual-branch DEN-KAN outperforms the single-path WA-KAN encoder across all seven crystal systems, with gains that scale inversely with symmetry. As shown in **Fig. 4a**, the largest improvement is 16.53% for the triclinic system, while hexagonal and cubic systems show smaller gains of 7.04% and 7.47%. These trends indicate that the SAXRD-aware design is most useful when high-angle fingerprints are less distinctive.

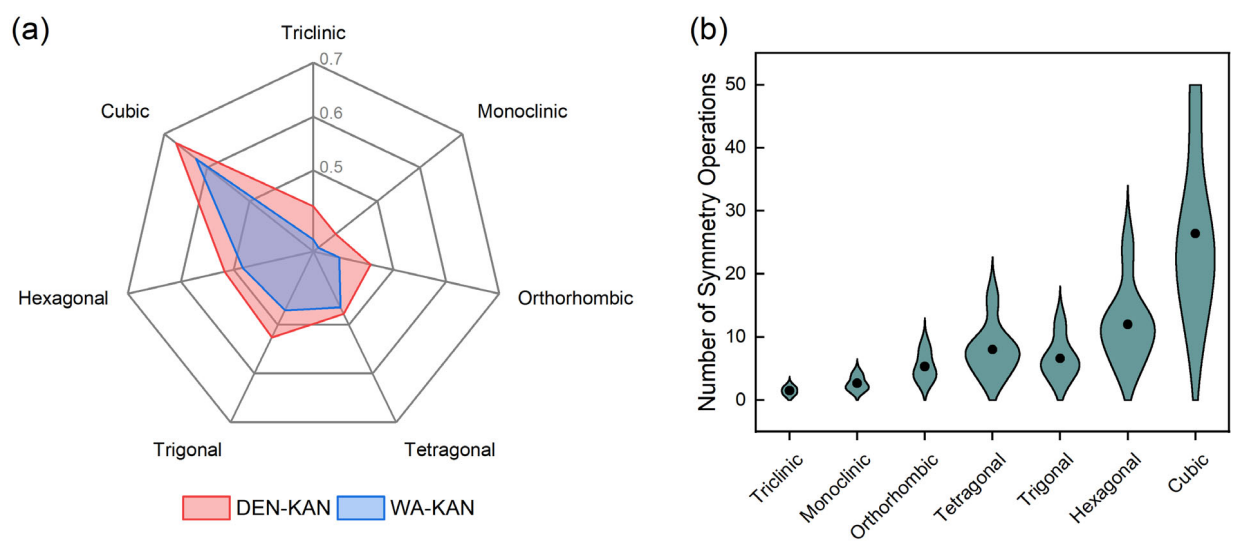

**Fig. 4. Encoder interpretability**. (a) Accuracy gain of DEN-KAN encoder over WA-KAN encoder and (b) symmetry operations across seven crystal systems.

The emphasis here is on the physical role of SAXRD. Small-angle measurements record reflections with large *d*-spacings by Bragg's law. Such peaks encode interlayer distances, staging, superlattice repeat lengths, guest-induced expansion, and other long-period motifs [54]. These signals are often weak yet highly discriminative when symmetry is low. Incorporating SAXRD therefore introduces complementary long-range information that is not accessible in WAXRD alone. The largest class-dependent gains appear where long-period features best separate candidates with similar high-angle patterns.

The fusion mechanism is governed by the KAN head, which acts as a physically informed aggregator rather than the primary accuracy driver. Adaptive spline activations combine SAXRD and WAXRD features while allowing local basis functions to represent sharp peaks and smooth backgrounds in a unified embedding. The head increases the contribution of SAXRD cues when high-angle information is less discriminative and reduces it when symmetry is high and WAXRD already provides rich fingerprints [54, 55]. The resulting embeddings respect crystallographic priors and remain stable during training. The system-specific improvements in **Fig. 4a** follow the distribution of symmetry operations in **Fig. 4b**, which supports the physical interpretability of the learned representation.

### 3.3 Application I: space group identification with XCCP

The space group encodes the discrete symmetry of a crystal and governs systematic absences, reflection multiplicities, and selection rules in powder diffraction. Because the XCCP framework is trained to align diffraction patterns with crystal structures, its ability to recover symmetry labels is a stringent test of physical fidelity and practical utility. Baselines for space group identification on the present MP-SXRD dataset cover a range of model families. The suite includes a fully convolutional neural network (FCN)[37], a residual network with an MLP classification head (ResNet-MLP)[53], a vision transformer (ViT)[56] that uses self-attention to capture long-range dependencies, and a hybrid that combine ResNet and ViT with an MLP head (ResViT-MLP). KAN classification heads were also tested, replacing MLP heads in ResNet and ResViT to form ResNet-KAN and ResViT-KAN. All models were trained and evaluated on intensity-normalized XRD profiles using the same data spit and training protocol. In XCCP, the space group of the top-1 retrieved structure was assigned as the predicted label for each pattern and compared against the ground truth.

**Table 1**. Space group prediction accuracy across different architectures without elemental information

| Model | Accuracy (SAXRD-aware) | Accuracy (WAXRD-only) |
|---|---|---|
| FCN | 48.60% | 48.05% |
| ViT | 49.65% | 49.15% |
| ResNet-MLP | 58.38% | 57.82% |
| ResViT-MLP | 59.54% | 58.74% |
| ResNet-KAN | 59.46% | 58.33% |
| ResViT-KAN | 59.75% | 59.47% |
| XCCP | 60.85% | 59.66% |

Most prior reports focus on WAXRD patterns. Representative results include 49.8% accuracy for a bidirectional gated recurrent unit on the SIMXRD-4M dataset [33] and 37.48% for an AutoML pipeline on the SIMPOD dataset[57]. These studies highlight the difficulty of the task under multi-scale features and motivates stronger XRD encoders with retrieval-aware predictors. Under the WAXRD-only condition in MP-SXRD, accuracies are 48.05% for FCN, 49.15% for ViT, 57.82% for ResNet-MLP, 58.74% for ResViT-MLP, 58.33% for ResNet-KAN, and 59.47% for ResViT-KAN. XCCP reaches 59.66% in the same setting, which places it at the top of this group. The ranking in **Table 1** indicates that residual learning and self-attention improve performance over plain convolutional stacks, and that geometry-aware heads such as KAN further narrow the gap to the best retrieval-guided predictor.

When SAXRD information is available, performance improves across all architectures. Accuracies become 48.60% for FCN, 49.65% for ViT, 58.38% for ResNet-MLP, 59.54% for ResViT-MLP, 59.46% for ResNet-KAN, and 59.75% for ResViT-KAN. XCCP benefits from the added small-angle cues and reaches 60.85%, an absolute gain of 1.19% over its WAXRD-only result. The improvement indicates that small-angle reflections provide complementary long-period descriptors that refine class boundaries, while the retrieval-aware design maintains decision consistency. For fairness, the comparisons above do not introduce elemental priors for XCCP. When elemental information is incorporated through the retrieval pipeline, accuracy in the SAXRD-aware setting rises sharply to 93.39%, which demonstrates the value of principled composition constraints.

**3.4. Application II: robustness on simulated and experimental datasets**

Beyond symmetry inference, practical deployment requires robustness across composition and data domains, and two scenarios thus are considered. The first involves multi-principal element alloys (MPEAs), where elemental similarity weakens composition cues. The second evaluates an experiment dataset from the opXRD database[52], where measurement variability is present. Elemental filtering is omitted for MPEAs because composition alone does not reliably distinguish phase in these systems. Elemental information is included for the opXRD set to match standard laboratory practice.

Phase identification in MPEAs presents unique challenges[14]. Small compositional changes can induce local distortions or symmetry-breaking, complicating diffraction profiles. FeCrAl-based and TaNbMo-based MPEA were selected as representative systems, both adopting BCC structures. Crystal structures were generated using the special quasi-random structure (SQS) method[58], and supercells were constructed with the Alloy Theoretic Automated Toolkit (ATAT) software[59]. **Fig. 5** shows representative crystal structures and their simulated XRD patterns. For alloys such as $Fe_{20}Cr_{10}Al_2$ and $Fe_{20}Cr_9Al_3$, only slight peak shifts and intensity changes are observed. A similar effect is seen for $Ta_{14}Nb_{14}Mo_{14}V_{12}$ and $Ta_{15}Nb_8Mo_{15}V_{16}$, where patterns are almost identical although atomic arrangements differ. Despite these difficulties, the XCCP framework achieves 66.67% top-1 accuracy and 95.87% top-3 accuracy across 22 crystal structures, as detailed in the **Supplementary Materials**. The framework successfully retrieves the correct structure even when peaks overlap due to subtle composition changes.

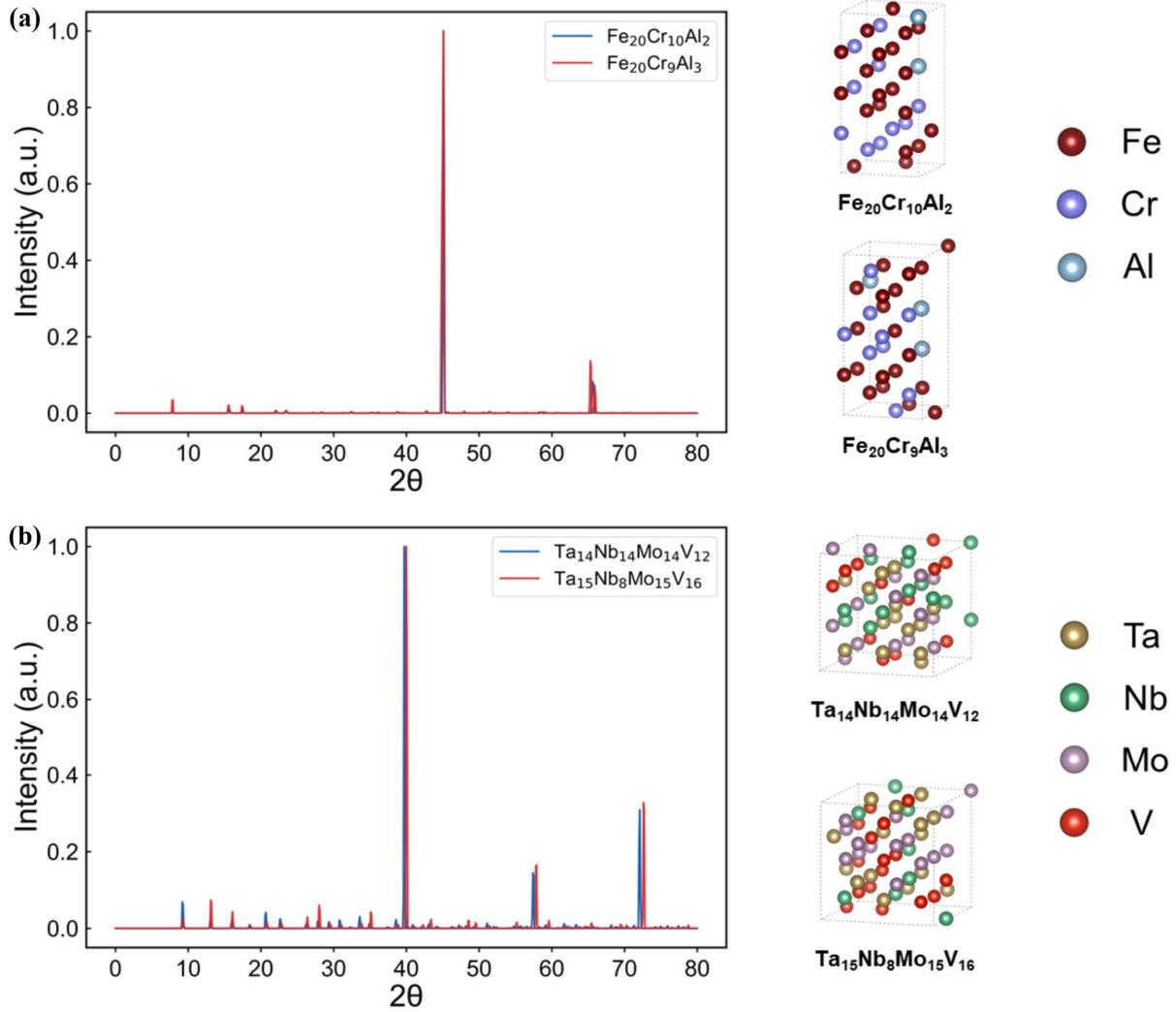

**Fig. 5.** Representative crystal structures and simulated patterns for (a) FeCrAl-based and (b) TaNbMo-based MPEAs.

A broader zero-shot evaluation on 773 experimental XRD patterns form the HKUST-subset of opXRD database further tests generality [52]. The experimental files lack reliable small-angle coverage, therefore the single-path WA-KAN variant is employed to match the available 2θ range. As summarized in **Table 2**, the XCCP framework demonstrates strong retrieval capabilities on the real-world dataset. The top-1 accuracy reaches 56.14%, and retrieval performance improves significantly as the retrieval scope expands. Top-3, top-5, and top-10 accuracies reach 84.61%, 93.40%, and 99.74%, respectively. The exceptionally high top-10 accuracy indicates that the correct structure almost always appears within ten candidates, which supports short-list inspection in experimental workflows. The ranking behavior aligns with the ablation results, where WA-KAN shows strong recall at larger $k$, and the absence of SAXRD primarily reduces early-rank precision rather than broad retrieval coverage.

Table 2.    Top-*k* accuracy of XCCP on experimental material database across seven crystal systems.

| Crystal system | top-1 | top-3 | top-5 | top-10 | Data Count |
|---|---|---|---|---|---|
| Triclinic | 60.27% | 85.48% | 94.16% | 99.67% | 599 |
| Monoclinic | 46.81% | 81.91% | 90.43% | 98.94% | 94 |
| Orthorhombic | 26.32% | 84.21% | 89.47% | 100.00% | 19 |
| Tetragonal | 33.33% | 50.00% | 66.67% | 100.00% | 6 |
| Trigonal | 45.95% | 78.38% | 97.30% | 100.00% | 37 |
| Hexagonal | 58.33% | 100.00% | 100.00% | 100.00% | 12 |
| Cubic | 100.00% | 100.00% | 100.00% | 100.00% | 6 |
| **Average** | **57.18%** | **84.73%** | **93.66%** | **99.61%** | **773** |

## 4. Discussion

This study introduces XCCP, a physics-guided contrastive framework that retrieves crystal structures directly from powder diffraction patterns. The dual-expert XRD encoder with a KAN projection head preserves long-range reflections, captures symmetry-driven peaks, and fuses them into interpretable, physically grounded embeddings. The approach performs well with the standard WAXRD profiles used in most laboratories and gains further early-rank precision when SAXRD information is available. These behaviors hold on individual simulations and carry over to experimental patterns.

Model ablations indicate that architectural choices matter. The WA and SA dual-branch encoder improves the quality of the diffraction representation, while the KAN head provides a decisive boost by aligning peak shapes and backgrounds with flexible local bases. As a result, retrieval accuracy and space-group recognition are competitive without elemental priors and become markedly stronger when composition constraints are applied.

The framework is readily extensible. Additional experimental evidence such as electron diffraction from transmission electron microscopy and X-ray scattering can be incorporated by adding new branches that are fused through the same KAN-based head. In addition, the embeddings after retrieval and similarity comparison can serve as inputs to downstream tasks including structure proposal, completion, or refinement under physical constraints. XCCP is a practical route to accurate, scalable, and interpretable crystallography. The method aligns with routine data acquisition, benefits from physics-informed design, and offers a clear path to multi-modality integration and generative pipelines that accelerate materials discovery.

## Author Contribution

**C. Xu** performed model training and drafted the manuscript. **S. Hao** contributed to data preprocessing and statistical analysis, and revised the manuscript. **Y. Wu** participated in the model design and training. **J. Xiong** conceived the research idea, performed data analysis, reviewed and revised the manuscript, and supervised the project. **S. Chen**, **S. Dong**, and **T. Jiang** participated in the interpretation and discussion of the results. **M. He** analyzed the results, co-conceived the research idea, and revised the manuscript. **T.Y. Zhang** reviewed the manuscript, and provided overall supervision.

## Data availability statement

Data will be made available on request

## Funding statement

This work was financially supported by the Advanced Materials-National Science and Technology Major Project (Grant No. 2025ZD0620102), National Natural Science Foundation of China (Grant No. 52401015), Shanghai Pujiang Program (Grant No. 23PJ1403500) and Shanghai Artificial Intelligence Open Source Award Project.